\documentclass[11pt]{article}
\usepackage{graphicx}

\begin{document}

\title{Mean Field Model of\\ Genetic Regulatory Networks}

\author{M. Andrecut, S. A. Kauffman}

\date{ }

\maketitle

{\par\centering Institute for Biocomplexity and Informatics \par}
{\par\centering University of Calgary \par}
{\par\centering 2500 University Drive NW, Calgary \par}
{\par\centering Alberta, T2N 1N4, Canada \par}

\noindent

\begin{abstract}
In this paper, we propose a mean-field model which attempts to bridge the
gap between random Boolean networks and more realistic stochastic modeling
of genetic regulatory networks. The main idea of the model is to replace all
regulatory interactions to any one gene with an average or effective
interaction, which takes into account the repression and activation
mechanisms. We find that depending on the set of regulatory parameters, the
model exhibits rich nonlinear dynamics. The model also provides quantitative
support to the earlier qualitative results obtained for random Boolean
networks.
\bigskip


\end{abstract}

\bigskip
\bigskip

\newpage

\section{Introduction}

Since its proposal, the Random Boolean Network model [1] has successfully
described in a qualitative way several important aspects of gene regulation
and cell differentiation processes (for references see [2]). The model is
constructed by assigning to each of the genes its regulatory inputs from
among the large number of genes present in the network. The model consists
of $N$ binary variables, corresponding to the two states of gene expression
(off and on). In this binary setting, each gene is assigned a logical
function on its inputs showing its next activity. While clearly an
idealization, much has been learned from this class of large Boolean
networks, and major features generalize to a class of piecewise linear
differential equations [3-4] and a family of polynomial maps [5]. The
research on complex Boolean networks, shows that networks behave in three
regimes: ordered, critical and chaotic [6]. It is a very attractive
hypothesis that cell types have evolved by natural selection to lie in the
ordered regime, close to the critical phase transition, where the most
complex coordinated behaviors can occur [7-8]. In this deterministic
setting, it is almost an inevitable hypothesis that the distinct cell types
of an organism correspond to the distinct attractors of the network. This
hypothesis needs to be confirmed by experimental tests. Cell differentiation
consists in response to a perturbation or signal that places the network in
a different basin of attraction from which it flows to a new attractor.
Obviously, real genetic networks are not Boolean nets. In real nets, one has
to take into account the molecular dynamics by using stochastic differential
equation models. The resulted equations can be solved using Monte-Carlo
methods. Here the favored approach is the Gillespie algorithm [9], which can
be used to model discrete molecular events of transcription, translation and
gene control in complex reaction networks. Boolean networks already impose
some formidable computational problems. Introducing stochastic models render
even smaller networks computationally intractable because the number of
reactions one has to consider grows exponentially fast with the number of
genes in the network.

Here, we propose a simplified mean-field model of the genetic regulatory
network, where the main idea is to replace all regulatory interactions to
any one gene with an average or effective interaction, which takes into
account the repression and activation mechanisms. Our model attempts to
bridge the gap between random Boolean networks and more realistic stochastic
modeling of regulatory networks. In this model, the regulatory interactions
are described by differential equations corresponding to the chemical
reactions considered in the genetic network. The same set of chemical
reactions can, for example, be used in a stochastic simulation of the
network. From this point of view, the proposed model gives a mean-field
description of the more accurate stochastic approach. In the mean-field
deterministic description, the gene-expression state at a given time and the
regulatory interactions among them unambiguously determine the
gene-expression state at the next time. In a stochastic system, on the other
hand, a given gene-expression state can generate more than one successive
gene-expression states, and therefore, different cells of the same
population may follow a different gene-expression path. As a result of these
considerations, the stochastic model describe the kinetics of gene
regulation more accurately than a deterministic model. However, a
deterministic mean-field model can be transformed in a stochastic model by
incorporating noise. This approach results in a stochastic differential
equation or Langevin equation. It is well known that the Langevin equation
is asymptotically equivalent (under certain conditions) to the chemical
master equation [10]. Therefore, the proposed mean-field model is still
relevant for the description of gene regulation and cell differentiation
processes. We show that depending on the set of regulatory parameters, the
model exhibits differing behaviors corresponding to ordered and chaotic
dynamics. This result gives quantitative support to the earlier qualitative
results obtained for random Boolean networks. Also, we show that the system
acquires stability by increasing the number of interactions. This conclusion
provides a possible explanation of how diversity and stability are created
in a biological system, giving rise to a great variety of stable living
organisms.

\section{The gene expression process}

For the beginning, let us analyze the gene expression process [11-14]. In a
genetic regulatory network, genes can be turned on or off by the binding of
proteins to regulatory sites on the DNA [1]. The proteins are known as
transcription factors, while the DNA binding sites are known as promoters.
Transcription factors can regulate the production of other transcription
factors, or they can regulate their own production. The simplest model of
gene expression involves only two steps in which the genetic information is
first transcribed into messenger RNA ($mRNA$) and then translated into
proteins ($M$) by ribosomes ($Ribo$). The transcription process can be
described by a sequence of reactions, in which the RNA polymerase ($RNAp$)
binds to the gene promoter ($P$) leading to transcription of a complete $mRNA
$ molecule: 
\begin{equation}
RNAp+P\stackrel{k_{1}}{\longrightarrow }C_{1}\stackrel{k_{2}}{
\longrightarrow }...\stackrel{k_{n}}{\longrightarrow }C_{n}\stackrel{k_{n+1}
}{\longrightarrow }RNAp+P+mRNA.
\end{equation}
Here, $C_{i}$ corresponds to the complex formed in the intermediate reaction 
$i=1,...,n$, with constant rate $k_{i}$.

Since the waiting times are independent statistical quantities, the waiting
time for the whole sequence of intermediate complex formation is the sum of
the waiting times for the individual steps. Also, we should note that the
central limit theorem [10] indicates that the lumped reaction of the open
complex formation will tend to have a Gaussian distribution of waiting
times, converging to a $\delta $ function for a very large number of
intermediate steps. Thus, in terms of reaction rates (which have units of
inverse time) we have $k^{-1}=\sum_{i=1}^{n+1}k_{i}^{-1}$.

From the above considerations, it follows that the whole sequence of
reactions can be approximated by the following reaction: 
\begin{equation}
RNAp+P\stackrel{k_{I}}{\longrightarrow }RNAp+P+mRNA,
\end{equation}
where 
\begin{equation}
k_{I}=\left( \sum_{i=1}^{n+1}\frac{1}{k_{i}}\right) ^{-1}.
\end{equation}

Let us now analyze the translation process, in which the information
initially transcribed into $mRNA$ is now translated into proteins $M$. To
describe this we consider the following additional reactions: 
\begin{equation}
Ribo+mRNA\stackrel{k_{II}}{\longrightarrow }Ribo+mRNA+M,
\end{equation}
\begin{equation}
mRNA\stackrel{\widetilde{k}_{II}}{\longrightarrow }\emptyset .
\end{equation}
The first reaction idealizes the multistep translation process, under the
further idealization that a ribosome ($Ribo$) binds the $mRNA$ and a protein 
$M$ is produced. The second reaction captures the degradation of $mRNA$.

\section{The mean-field model}

The model described here is intentionally as simple as possible. We believe
that such an approach is as important as detailed biological modeling in
elucidating the basic physics behind genetic regulatory networks. The main
idea of the model is to replace all regulatory interactions to any one gene
with an average or effective interaction which takes into account the
repression and activation mechanisms.

Let us now consider a system formed from $N$ genes. In this system protein
multimers (which are the transcription factors) are responsible for gene
regulation (activation and repression) and are allowed to bind to the
promoter site. Here, an important problem is to specify which reactions are
allowed to take place in the system. We consider that the basic level of
expression (transcrition and translation), the $mRNA$ and protein
degradation reactions always exist, and they represent the minimum set of
reactions describing the system. The existence of the other (activation,
repression and multimerization) reactions is specified by an associated set
of binary coefficients (switches) $\alpha $, $\beta $, $\gamma $, $\lambda $
, $\mu \in \{0,1\}$. If the value of such a coefficient is $1$ then the
reaction exists, if the coefficient is $0$ then the reaction does not exist.
The possible chemical reactions up to the dimer interaction case are $
(n,i,j=1,...,N)$:

- Transcription: 
\begin{equation}
RNAp+P_{n}\stackrel{k_{I}^{n}}{\longrightarrow }RNAp+P_{n}+mRNA_{n},
\end{equation}

- $mRNA$ degradation 
\begin{equation}
mRNA_{n}\stackrel{\widetilde{k}_{I}^{n}}{\longrightarrow }\emptyset .
\end{equation}

- Translation:
\begin{equation}
Ribo+mRNA_{n}\stackrel{k_{II}^{n}}{\longrightarrow }Ribo+mRNA_{n}+M_{n}.
\end{equation}

- Protein degradation: 
\begin{equation}
M_{n}\stackrel{\widetilde{k}_{II}^{n}}{\longrightarrow }\emptyset .
\end{equation}

- Protein dimerization: 
\begin{equation}
\alpha _{ij}:\quad M_{i}+M_{j}\stackrel{\widetilde{k}_{ij}}{\longleftarrow }
\stackrel{k_{ij}}{\longrightarrow }M_{i}M_{j},
\end{equation}
where the coefficients $\alpha _{ij}=\alpha _{ij}\in \{0,1\}$ are the
selection switches. Here, a protein dimer is formed by combining two protein
monomers.

- Repression: 
\begin{equation}
\beta _{i}^{n}:\quad P_{n}+M_{i}\stackrel{\ \widetilde{k}_{i}^{n}}{
\longleftarrow }\stackrel{k_{i}^{n}}{\longrightarrow }P_{n}M_{i},
\end{equation}
\begin{equation}
\lambda _{ij}^{n}\alpha _{ij}:\quad P_{n}+M_{i}M_{j}\stackrel{\widetilde{k}
_{ij}^{n}}{\longleftarrow }\stackrel{k_{ij}^{n}}{\longrightarrow }
P_{n}M_{i}M_{j},
\end{equation}
where the selection is done by setting $\beta _{i}^{n}$, $\lambda
_{ij}^{n}=\lambda _{ji}^{n}\in \{0,1\}$. In the first reaction, a protein
monomer binds to the promoter and the effect is an inhibition of the basic
level of transcription reaction, which actually leads to a gene repression.
The existence of the second repression reaction is conditioned by the
dimerization reaction between proteins. Therefore, the real regulatory
coefficient for this reaction is $\lambda _{ij}^{n}\alpha _{ij}$.

- Activation: 
\begin{equation}
\gamma _{i}^{n}\beta _{i}^{n}:\quad RNAp+P_{n}M_{i}\stackrel{k_{i}^{n}}{
\longrightarrow }RNAp+P_{n}M_{i}+mRNA_{n},
\end{equation}
\begin{equation}
\mu _{ij}^{n}\lambda _{ij}^{n}\alpha _{ij}:\quad RNAp+P_{n}M_{i}M_{j}
\stackrel{k_{ij}^{n}}{\longrightarrow }RNAp+P_{n}M_{i}M_{j}+mRNA_{n}.
\end{equation}
In order to have an activation, a protein monomer or a dimer must first bind
to the promoter and form another complex. Therefore, the existence of an
activation reaction is conditioned by the coefficients $\beta _{i}^{n}$, $
\lambda _{ij}^{n}$ and $\alpha _{ij}$. Thus, the activation reactions are in
fact regulated by $\gamma _{i}^{n}\beta _{i}^{n}$, $\mu _{ij}^{n}\lambda
_{ij}^{n}\alpha _{ij}$, where $\gamma _{i}^{n}$, $\mu _{ij}^{n}=\mu
_{ij}^{n}\in \{0,1\}$.

We should note that the number of possible reactions is very high. This
number depends on the considered length of possible multimers formed by the
transcription factors. Here we have considered only the reactions up to a
possible dimer interaction. However, we will show that these reactions are
enough for the purpose of illustrating the interaction mechanism in a more
general model, corresponding to higher order multimer interactions.

In a steady state the repression and dimerization reactions are in
equilibrium and we have: 
\begin{eqnarray}
k_{i}^{n}[P_{n}][M_{i}] &=&\widetilde{k}_{i}^{n}[P_{n}M_{i}], \\
k_{ij}^{n}[P_{n}][M_{i}M_{j}] &=&\widetilde{k}_{ij}^{n}[P_{n}M_{i}M_{j}], 
\nonumber \\
k_{ij}[M_{i}][M_{j}] &=&\widetilde{k}_{ij}[M_{i}M_{j}].  \nonumber
\end{eqnarray}
Here by $[A]$ we understand the concentration of $A$. One can see that each
promoter $P_{n}$ can be in different states corresponding to the
transcription factor complex which binds to it: 
\begin{equation}
\{P_{n},P_{n}M_{i},P_{n}M_{i}M_{j}\}.
\end{equation}
From the above equations one can see that the probabilities associated to
these states are in the ratio: 
\begin{eqnarray}
\frac{\lbrack P_{n}M_{i}]}{[P_{n}]} &=&\frac{k_{i}^{n}}{\widetilde{k}_{i}^{n}
}[M_{i}]=a_{i}^{n}x_{i}, \\
\frac{\lbrack P_{n}M_{i}M_{j}]}{[P_{n}]} &=&\frac{k_{ij}^{n}}{\widetilde{k}
_{ij}^{n}}[M_{i}M_{j}]=\frac{k_{ij}^{n}}{\widetilde{k}_{ij}^{n}}\frac{k_{ij}
}{\widetilde{k}_{ij}}x_{i}x_{j}  \nonumber \\
&=&(b_{ij}^{n}+b_{ij}^{n})x_{i}x_{j}=2b_{ij}^{n}x_{i}x_{j}.  \nonumber
\end{eqnarray}
where $x_{i}$ are the reduced concentrations variables for the transcription
factors.

The probability that the promoter $P_{n}$ is in a free state is given by: 
\begin{equation}
\pi _{0}^{n}=\frac{1}{1+\sum\limits_{i}\beta
_{i}^{n}a_{i}^{n}x_{i}+\sum\limits_{ij}\lambda _{ij}^{n}\alpha
_{ij}b_{ij}^{n}x_{i}x_{j}},
\end{equation}
where the denominator is the sum over all possible states of the promoter
including the free state and the binding states. Consequently, the
probabilities that the promoter is in a binding state $P_{n}M_{i}$ or $
P_{n}M_{i}M_{j}$ are: 
\begin{equation}
\pi _{i}^{n}=\frac{\beta _{i}^{n}a_{i}^{n}x_{i}}{1+\sum\limits_{i}\beta
_{i}^{n}a_{i}^{n}x_{i}+\sum\limits_{ij}\lambda _{ij}^{n}\alpha
_{ij}b_{ij}^{n}x_{i}x_{j}},
\end{equation}
\begin{equation}
\pi _{ij}^{n}=\frac{\lambda _{ij}^{n}\alpha _{ij}b_{ij}^{n}x_{i}x_{j}}{
1+\sum\limits_{i}\beta _{i}^{n}a_{i}^{n}x_{i}+\sum\limits_{ij}\lambda
_{ij}^{n}\alpha _{ij}b_{ij}^{n}x_{i}x_{j}}.
\end{equation}
Obviously, from the above equations we have: 
\begin{equation}
\pi _{0}^{n}+\sum_{i}\pi _{i}^{n}+\sum_{ij}\pi _{ij}^{n}=1.
\end{equation}
In a steady state, the rate of $mRNA_{n}$ transcription should be equal to
the rate of $mRNA_{n}$ degradation and therefore we can write the following
mean-field equation: 
\begin{equation}
\frac{k_{I}^{n}+\sum\limits_{i}k_{i}^{n}\gamma _{i}^{n}\beta
_{i}^{n}a_{i}^{n}x_{i}+\sum\limits_{i,j}k_{ij}^{n}\mu _{ij}^{n}\lambda
_{ij}^{n}\alpha _{ij}b_{ij}^{n}x_{i}x_{j}}{1+\sum\limits_{i}\beta
_{i}^{n}a_{i}^{n}x_{i}+\sum\limits_{ij}\lambda _{ij}^{n}\alpha
_{ij}b_{ij}^{n}x_{i}x_{j}}=\widetilde{k}_{I}^{n}y_{n},
\end{equation}
where $y_{n}$ is the reduced concentration of $mRNA_{n}$. Here, we have
covered all the system construction possibilities by using the binary
coefficients $\alpha _{ij}$, $\beta _{i}^{n}$, $\gamma _{i}^{n}$, $\lambda
_{ij}^{n}$, $\mu _{ij}^{n}$. Thus, the transcription of gene $n$, can be
approximated using the following nonlinear differential equation: 
\begin{equation}
\frac{d}{dt}y_{n}=\eta _{n}\frac{1+\sum\limits_{i}\gamma _{i}^{n}\beta
_{i}^{n}c_{i}^{n}a_{i}^{n}x_{i}+\sum\limits_{ij}\mu _{ij}^{n}\lambda
_{ij}^{n}\alpha _{ij}d_{ij}^{n}b_{ij}^{n}x_{i}x_{j}}{1+\sum\limits_{i}\beta
_{i}^{n}a_{i}^{n}x_{i}+\sum\limits_{ij}\lambda _{ij}^{n}\alpha
_{ij}b_{ij}^{n}x_{i}x_{j}}-y_{n},
\end{equation}
where $c_{i}^{n}=k_{i}^{n}/k_{I}^{n}$, $d_{ij}^{n}=k_{ij}^{n}/k_{I}^{n}$.
Also, $\eta _{n}=k_{I}^{n}/\widetilde{k}_{I}^{n}$ is a parameter
corresponding to the ''promoter strength'' (the ratio between the
transcription and the degradation rates).

Also, in a steady state, the rate of translation should be equal to the rate
of protein degradation. Therefore, for the gene $n$, we can write the second
equation: 
\begin{equation}
k_{II}^{n}[mRNA_{n}]=\widetilde{k}_{II}^{n}[M_{n}],
\end{equation}
and consequently the second differential equation: 
\begin{equation}
\frac{d}{dt}x_{n}=\theta _{n}(y_{n}-x_{n}),
\end{equation}
where $\theta _{n}=k_{II}^{n}/\widetilde{k}_{II}^{n}$ is a coefficient
measuring the delay induced by the translation process.

Let us give some simple examples to the above equations. First, let us
consider the case of two genes, where all the monomer interactions are
excluded and the interaction is mediated only by dimers ($\beta
_{i}^{n}=\gamma _{i}^{n}=0$). We consider that the interaction at the dimer
level is performed only by homo-dimers ($\alpha _{ij}=0,i\neq j)$. Also, the
interaction is due only to the repression mechanism ($\mu _{ij}^{n}=0$), and
the repression is performed only by the other gene's homo-dimers ($\lambda
_{11}^{1}=\lambda _{22}^{1}=\lambda _{12}^{1,2}=0$). With these
approximations we obtain the well known toggle switch model [15]: 
\begin{eqnarray}
\frac{d}{dt}y_{1} &=&\frac{\eta _{1}}{1+b_{22}^{1}x_{2}^{2}}-y_{1}, \\
\frac{d}{dt}y_{2} &=&\frac{\eta _{2}}{1+b_{11}^{2}x_{1}^{2}}-y_{2}, 
\nonumber \\
\frac{d}{dt}x_{1} &=&\theta _{1}(y_{1}-x_{1}),  \nonumber \\
\frac{d}{dt}x_{2} &=&\theta _{2}(y_{2}-x_{2})  \nonumber
\end{eqnarray}
Now, let us consider the case in which three genes are repressing each other
at a dimer level in a circular way: $1\leftarrow 2\leftarrow 3\leftarrow
1\leftarrow 2\leftarrow 3...$ This means that: $\beta _{i}^{n}=\gamma
_{i}^{n}=0$; $\alpha _{ij}=0$ ($i\neq j$); $\mu _{ij}^{n}=0$; $\lambda
_{ij}^{n}=0$ ($i\neq j$ and $i=j\neq mod(3)+1$). With these
assumptions we obtain the well known equations of the repressilator [16]: 
\begin{eqnarray}
\frac{d}{dt}y_{1} &=&\frac{\eta _{1}}{1+b_{22}^{1}x_{2}^{2}}-y_{1}, \\
\frac{d}{dt}y_{2} &=&\frac{\eta _{2}}{1+b_{33}^{2}x_{3}^{2}}-y_{2}, 
\nonumber \\
\frac{d}{dt}y_{3} &=&\frac{\eta _{3}}{1+b_{11}^{2}x_{1}^{2}}-y_{3}, 
\nonumber \\
\frac{d}{dt}x_{1} &=&\theta _{1}(y_{1}-x_{1}),  \nonumber \\
\frac{d}{dt}x_{2} &=&\theta _{1}(y_{2}-x_{2}),  \nonumber \\
\frac{d}{dt}x_{3} &=&\theta _{1}(y_{3}-x_{3}).  \nonumber
\end{eqnarray}
The analysis of the repressilator model has shown that in order to obtain
sustained oscillations we need strong promoters, which is equivalent to a
high rate of expression or a low rate of degradation [16].

From the above considerations we observe that the multimer interactions can
be represented using a tensorial expansion. The coefficients describing the
interaction can be condensed in tensors of different ranks, corresponding to
the length of considered multimers. Thus, in general for a network with $N$
genes we can write the following set of equations $(i,j,n=1,...,N)$: 
\begin{eqnarray}
\frac{d}{dt}y_{n} &=&\eta _{n}\frac{1+\sum\limits_{i}\gamma _{i}^{n}\beta
_{i}^{n}c_{i}^{n}a_{i}^{n}x_{i}+\sum\limits_{i,j}\mu _{ij}^{n}\lambda
_{ij}^{n}\alpha _{ij}d_{ij}^{n}b_{ij}^{n}x_{i}x_{j}+...}{1+\sum\limits_{i}
\beta _{i}^{n}a_{i}^{n}x_{i}+\sum\limits_{ij}\lambda _{ij}^{n}\alpha
_{ij}b_{ij}^{n}x_{i}x_{j}+...}-y_{n}, \\
\frac{d}{dt}x_{n} &=&\theta _{n}(y_{n}-x_{n}).  \nonumber
\end{eqnarray}
One can see that the numerator of the above fraction contains all the
activation interactions, while the denominator contains all the repression
interactions. Therefore, this intrinsic ratio, between activation and
repression, defines the dynamics of the gene network.

\section{Numerical results}

In order to simplify further our description we consider that the
transcription and translation processes can be condensed in only one
reaction: 
\begin{equation}
RNAp+P\stackrel{k}{\longrightarrow }RNAp+P+M.
\end{equation}
This means that we are neglecting the delay effects introduced by the
translation process, which is equivalent to consider $y_{n}=x_{n}$.
Therefore the dynamics of the system will be described only by the following
simplified equations: 
\begin{equation}
\frac{d}{dt}x_{n}=\eta _{n}\frac{1+\sum\limits_{i}\gamma _{i}^{n}\beta
_{i}^{n}c_{i}^{n}a_{i}^{n}x_{i}+\sum\limits_{i,j}\mu _{ij}^{n}\lambda
_{ij}^{n}\alpha _{ij}d_{ij}^{n}b_{ij}^{n}x_{i}x_{j}+...}{1+\sum\limits_{i}
\beta _{i}^{n}a_{i}^{n}x_{i}+\sum\limits_{ij}\lambda _{ij}^{n}\alpha
_{ij}b_{ij}^{n}x_{i}x_{j}+...}-x_{n}.
\end{equation}

Let us consider the mean-field equations up to the dimer level. Obviously,
these equations contain an extremely large number of unknown parameters. For
example, very few rate constants for the constituent reactions have been
measured in cells and one is often ignorant of absolute concentrations of
the participating molecular species. Meanwhile, the ensemble approach [17],
which consists in sampling random networks from an ensemble of networks
built according to the constraints we know that characterize real genomic
systems and then analyzing the typical, or generic, properties of ensemble
members, remains one useful approach to making use of the information we are
gathering on real systems to understand their large scale dynamical and
network connectivity implications.

We would like to emphasize that the mean-field model described here has the
advantage that all the parameters correspond only to ratios of reaction
constant rates, and therefore it does not require the knowledge of absolute
values of these constant rates. This characteristic of the model gives us
the chance to simplify even more the sampling of the parameters. Thus, in
the spirit of the ensemble approach let us consider that the coefficients $
\eta _{n}$, $a_{i}^{n}$, $c_{i}^{n}$, $b_{ij}^{n}$, $d_{ij}^{n}$, $
(i,j,n=1,...,N)$ are drawn from a Gaussian distribution as following: 
\begin{eqnarray}
\eta _{n} &=&\overline{\eta }(1+\sigma _{1}\xi ), \\
a_{i}^{n} &=&\overline{a}(1+\sigma _{2}\xi ),  \nonumber \\
c_{i}^{n} &=&\overline{c}(1+\sigma _{3}\xi ),  \nonumber \\
b_{ij}^{n} &=&\overline{b}(1+\sigma _{4}\xi ),  \nonumber \\
d_{ij}^{n} &=&\overline{d}(1+\sigma _{5}\xi ).  \nonumber
\end{eqnarray}
Here, $\xi $ is a random variable governed by a Gaussian distribution with
zero mean and variance equal to one, and $\sigma _{i}\geq 0$. By using
this sampling procedure, we assume implicitly that the ratios of reaction
constant rates are normally distributed around some average values $
\overline{\eta }$, $\overline{a}$, $\overline{c}$, $\overline{b}$, $
\overline{d}$. Also, high values of the parameters controlling the variance (
$\sigma _{i}$) are useful in creating a large spectrum of values around the
average values. For example, because $c_{i}^{n}=k_{i}^{n}/k_{I}^{n}$ are
generated from a Gaussian distribution centered at $\overline{c}$ it follows
that: $k_{i}^{n}\leq \overline{c}k_{I}^{n}$ or $k_{i}^{n}\geq \overline{c}
k_{I}^{n}$. This means that for $\overline{c}=1$ the rate of activation
reactions can be higher or lower than the rate corresponding to the basic
level of expression. In our simulations we have used the following
parameters: $\sigma _{i}=0.25$, $\overline{a}=\overline{c}=\overline{b}=
\overline{d}=1$, $\overline{\eta }=10^{5}$. Also, the initial conditions $
x_{n}(t=0)$ are drawn from an uniform distribution between $0$ and $\eta _{n}
$. Obviously, one can try a different scenario in which for example all
these parameters are drawn from an uniform distribution.

\subsection{M=1 (only monomers)}

Let us now consider the extreme case when all the dimers formed by the
transcription factors are missing from the system: 
\begin{equation}
s_{\alpha }=\sum\limits_{ij}\alpha _{ij}=0.
\end{equation}
Because the number of parameters in the system is still very large one can
imagine a huge number of numerical experiments. Below we describe a couple
of such possible numerical experiments.

In the first experiment, the coefficients regulating the monomer
interactions $\beta _{i}^{n},\gamma _{i}^{n}\in \{0,1\}$ are generated
randomly such that: 
\begin{eqnarray}
\beta _{i}^{n} &=&\left\{ 
\begin{array}{lll}
1 & with\ probability & p \\ 
0 & with\ probability & 1-p
\end{array}
\right. , \\
\gamma _{i}^{n} &=&\left\{ 
\begin{array}{lll}
1 & with\ probability & q \\ 
0 & with\ probability & 1-q
\end{array}
\right. .  \nonumber
\end{eqnarray}
The numerical results have shown that for any values of $p$ and $q$ the
system converges only to steady states.

In the second experiment, the values of the coefficients regulating the
monomer interactions $\beta _{i}^{n},\gamma _{i}^{n}\in \{0,1\}$ are
generated such that the sums: 
\begin{eqnarray}
s_{\beta }^{n} &=&\sum\limits_{i}\beta _{i}^{n}, \\
s_{\gamma }^{n} &=&\sum\limits_{i}\gamma _{i}^{n}.  \nonumber
\end{eqnarray}
follow a power law distribution: 
\begin{equation}
P(s)=\frac{1}{\zeta (\omega )}s^{-\omega },
\end{equation}
where $\zeta (\omega )=\sum_{s=1}^{\infty }s^{-\omega }$ is the Riemann Zeta
function. Recent analysis indicates that a power law distribution of
interactions seems to fit better the data observed experimentally for
several organisms [18-20]. Such a distribution can be obtained from an
uniform distribution using the inverse transformation method: 
\begin{equation}
s=[r(N_{\max }^{1-\omega }-N_{\min }^{1-\omega })+N_{\min }^{1-\omega }]^{
\frac{1}{1-\omega }}.
\end{equation}
Here, $r$ is an uniform distributed random number on the interval $[0,1]$.
The exponent of the power law distribution is $\omega =1+\delta $ (where $
\delta $ is the Pareto distribution shape parameter). The above equation
returns a value $s\in [N_{\min },N_{\max }]$, distributed accordingly to a
power law with the exponent $\omega $. In our simulation we have set $
N_{\min }=1$, $N_{\max }=N$ and the variable parameter is $\omega $. The
rest of the parameters and the initial states are set as before. The
numerical results have shown that for any values $\omega >1$ the system
converges to steady states. These numerical results suggest that protein
multimerization might be a necessary condition to generate more complex
dynamics in the system.

\subsection{M=2 (monomers and dimers)}

By increasing the number of dimers formed by the transcription factors: 
\begin{equation}
0<s_{\alpha }=\sum\limits_{ij}\alpha _{ij}<N^{2},
\end{equation}
one can easily see how the complex behavior emerges in the system. Depending
on the values of the other regulatory parameters, the model exhibits complex
oscillatory and chaotic dynamics.

Let us now consider the other extreme case when all the dimers formed by the
transcription factors are present in the system: 
\begin{equation}
s_{\alpha }=\sum\limits_{ij}\alpha _{ij}=N^{2}.
\end{equation}

The coefficients $\beta _{i}^{n},\gamma _{i}^{n}\in \{0,1\}$, $(i,n=1,...,N)$
, regulating the monomer interactions, are generated randomly as before,
with the probabilities $p$ and $q$. The other coefficients are determined as
following: 
\begin{eqnarray}
\lambda _{ij}^{n} &=&\left\{ 
\begin{array}{lll}
1 & with\ probability & p^{\prime } \\ 
0 & with\ probability & 1-p^{\prime }
\end{array}
\right. , \\
\mu _{ij}^{n} &=&\left\{ 
\begin{array}{lll}
1 & with\ probability & q^{\prime } \\ 
0 & with\ probability & 1-q^{\prime }
\end{array}
\right. .  \nonumber
\end{eqnarray}
The rest of the coefficients and the initial conditions are set as before in
the numerical experiment for $M=1$. The numerical results, have shown that
for any values of $p$, $q$, $p^{\prime }$, $q^{\prime }$ the system mostly
converges to steady states similar to those obtained for $M=1$, however
oscillatory solutions have also been observed.

In the next numerical experiment we consider that the values of the
coefficients regulating the monomer and dimer interactions are generated
such that the sums: 
\begin{eqnarray}
s_{\beta }^{n} &=&\sum\limits_{i}\beta _{i}^{n}, \\
s_{\gamma }^{n} &=&\sum\limits_{i}\gamma _{i}^{n},  \nonumber \\
s_{\lambda }^{n} &=&\sum\limits_{ij}\delta _{ij}^{n},  \nonumber \\
s_{\mu }^{n} &=&\sum\limits_{ij}\mu _{ij}^{n},  \nonumber
\end{eqnarray}
follows a power law distribution. For $s_{\beta }^{n}$ and $s_{\gamma }^{n}$
we set $N_{\max }=N$, while for $s_{\lambda }^{n}$ and $s_{\mu }^{n}$ we
have $N_{\max }=N^{2}$. The rest of the coefficients and the initial
conditions are set as before. For small values of $\omega \in (1,2)$ the
system converges mostly to steady states. An example of steady state
obtained for $\omega =1.5$ and $N=100$ is given in Fig. 1(a). For average
values $\omega \in [2,2.5)$ the density of steady states in the solution
space decreases, making room for oscillatory solutions. In Fig. 1(b) we give
an example of oscillatory solution obtained for $\omega =2.25$. For larger
values $\omega \geq 2.5$ one can easily obtain chaotic solutions. In Fig.
1(c) we give an example of chaotic solution obtained for $\omega =3$.

The dynamic behavior of solutions was characterized by performing Fourier
analysis on long trajectories. The Fourier power spectrum is discrete for an
oscillatory solution, while in the case of a chaotic solution it shows
continuous intervals of frequencies. In Fig. 2 we give the typical Fourier
power spectrum obtained for a single $x_{n}(t)$ trajectory: (a)
oscillations; (b) chaos.

The above result suggests that for power law distributed interactions (up to
the dimer level), there is a transition between order and chaos when the
power law exponent $\omega $ increases. In order to characterize this
transition one can try to calculate the largest Lyapunov exponent for the
dynamical system [21]. A positive value of the largest Lyapunov exponent
indicates the chaotic behavior of the system, while a negative value
indicates a regular dynamics. We have performed several calculations for
networks with a modest size ($N\sim 10$) just to confirm the chaotic
behavior of the solutions. Unfortunately, the computation of the Lyapunov
exponent is quite intensive and it quickly becomes prohibitive for networks
with a very large number of genes, sampled from a huge ensemble of networks
(as described above). Therefore, in order to analyze this transition we
consider a simplified approach, which focuses on steady state solutions, by
calculating the average fluctuations: 
\begin{equation}
\Omega =\sqrt{\frac{1}{NT}\sum\limits_{n=1}^{N}\sum
\limits_{t=1}^{T}(x_{n}(t)-\overline{x}_{n})^{2}}.
\end{equation}
Here, $T$ is the length of the trajectory (the number of time steps, after
the transient is eliminated) and $\overline{x}_{n}=T^{-1}\sum_{t}x_{n}(t)$
is the time average of $x_{n}(t)$. The global quantity $\Omega $ measures
the fluctuations around the average values $\overline{x}_{n}$. Obviously, $
\Omega $ cannot distinguish between chaotic and oscillatory trajectories,
however it provides a simple and efficient discrimination between ordinary
steady state solutions (Fig. 1(a)) and the oscillatory and chaotic solutions
(Fig. 1(b), (c)). For ordinary steady state solutions $\Omega \sim 0$, while
for oscillatory and chaotic solutions $\Omega \gg 0$. In Fig. 3. we give the
results obtained for $\Omega (\omega )$ by averaging over 100 solutions for $
\omega \in [1.5,4]$. The parameter $\Omega $ measures the transition between
a phase with high density of ordinary steady state solutions and a phase
with low density of steady state solutions. One can see that the transition
occurs around $\omega _{c}\simeq 2.25$. Also, one can see that $\omega _{c}$
does not seems to depend on the number of genes in the network $N=50$, $75$, 
$100$, $125$, $150$. Therefore, the average number of interactions per gene
at the critical point in a large network ($N\rightarrow \infty $) can be
easily estimated as: 
\begin{equation}
s_{c}=\left\langle s_{\beta /\gamma /\lambda /\mu }^{n}\right\rangle
_{\omega _{c}}=\frac{\zeta (\omega _{c}-1)}{\zeta (\omega _{c})}\simeq
3.146...
\end{equation}
Obviously, by increasing $\omega $ the number of interactions per gene
decreases. For large values of $\omega $ the number of interactions
approaches $s_{\beta /\gamma /\lambda /\mu }^{n}\sim 1$. These results
suggest that the network is more stable for low values of $\omega $, i.e.
when the number of interactions per gene is higher than $s_{c}$, and it
looses stability when $\omega $ increases, i.e. when the number of
interactions is low. We have observed this phenomenon for various values of
the system parameters, which suggests that it is an important characteristic
of the system.

\subsection{M$\geq $3 (multimers up to the rank M)}

For $M\geq 3$ the numerical simulation becomes difficult because of the
higher order combinatorial explosion in the definition of tensor
coefficients. The number of coefficients for a tensor with rank $M$ grows
exponentially fast as $N^{M}$, where $N$ is the number of genes in the
network. Therefore, further simplifications are required. Here, we consider
a simplified system of $N$ genes which has a high level of complexity,
comparable with the general mean-field system of equations. This simplified
system has the advantage that it can make the simulations possible even for
very large values of $M$ and $N$.

We make a further simplification by assuming that 
\begin{equation}
c_{i}^{n}a_{i}^{n}=c_{n}a_{n},\quad
d_{ij}^{n}b_{ij}^{n}=(c_{n}a_{n})^{2},\quad ...\quad (i,j,n=1,...,N).
\end{equation}
This means that the interaction strength at a given multimer level depends
slightly on the corresponding tensor rank. With these assumptions we obtain
the following set of equations ($i,j,n=1,...,N$): 
\begin{equation}
\frac{d}{dt}x_{n}=\eta _{n}\frac{1+a_{n}c_{n}\sum\limits_{i}\gamma
_{i}^{n}\beta _{i}^{n}x_{i}+(a_{n}c_{n})^{2}\sum\limits_{ij}\mu
_{ij}^{n}\lambda _{ij}^{n}\alpha _{ij}x_{i}x_{j}+...}{1+a_{n}\sum\limits_{i}
\beta _{i}^{n}x_{i}+a_{n}^{2}\sum\limits_{ij}\lambda _{ij}^{n}\alpha
_{ij}x_{i}x_{j}+...}-x_{n}.
\end{equation}

Now let us assume that all the higher order multimer interactions
(corresponding to a tensor rank $m\leq M$) are generated from the first
order monomer interactions in the following recursive way:

-repression: 
\begin{eqnarray}
\left( \sum\limits_{i}\beta _{i}^{n}x_{i}\right) ^{m}
&=&\sum\limits_{i_{1}}\beta _{i_{1}}^{n}x_{i_{1}}...\sum\limits_{i_{m}}\beta
_{i_{m}}^{n}x_{i_{m}} \\
&=&\sum\limits_{i_{1}}...\sum\limits_{i_{m}}\beta _{i_{1}}^{n}...\beta
_{i_{m}}^{n}x_{i_{1}}...x_{i_{m}};  \nonumber
\end{eqnarray}

-activation: 
\begin{eqnarray}
\left( \sum\limits_{i}\gamma _{i}^{n}\beta _{i}^{n}x_{i}\right) ^{m}
&=&\sum\limits_{i_{1}}\gamma _{i_{1}}^{n}\beta
_{i_{1}}^{n}x_{i_{1}}...\sum\limits_{i_{m}}\gamma _{i_{m}}^{n}\beta
_{i_{m}}^{n}x_{i_{m}} \\
&=&\sum\limits_{i_{1}}...\sum\limits_{i_{m}}\gamma _{i_{1}}^{n}\beta
_{i_{1}}^{n}...\gamma _{i_{m}}^{n}\beta _{i_{m}}^{n}x_{i_{1}}...x_{i_{m}}. 
\nonumber
\end{eqnarray}
For example, the dimer interaction terms are given by:

-repression: 
\begin{eqnarray}
\sum\limits_{ij}\lambda _{ij}^{n}\alpha _{ij}x_{i}x_{j}
&=&\sum\limits_{i}\sum\limits_{j}\beta _{i}^{n}\beta _{j}^{n}x_{i}x_{j} \\
&=&\sum\limits_{i}\beta _{i}^{n}x_{i}\sum\limits_{j}\beta _{j}^{n}x_{j} 
\nonumber \\
&=&\left( \sum\limits_{i}\beta _{i}^{n}x_{i}\right) ^{2};  \nonumber
\end{eqnarray}

-activation: 
\begin{eqnarray}
\sum\limits_{ij}\mu _{ij}^{n}\lambda _{ij}^{n}\alpha _{ij}x_{i}x_{j}
&=&\sum\limits_{i}\sum\limits_{j}\gamma _{i}^{n}\beta _{i}^{n}\gamma
_{j}^{n}\beta _{j}^{n}x_{i}x_{j} \\
&=&\sum\limits_{i}\gamma _{i}^{n}\beta _{i}^{n}x_{i}\sum\limits_{j}\gamma
_{j}^{n}\beta _{j}^{n}x_{j}  \nonumber \\
&=&\left( \sum\limits_{i}\gamma _{i}^{n}\beta _{i}^{n}x_{i}\right) ^{2}. 
\nonumber
\end{eqnarray}
Thus, we obtain the following simplified system of differential equations $
(i,n=1,...,N)$: 
\begin{equation}
\frac{d}{dt}x_{n}=\eta _{n}\left[ \frac{\left(
c_{n}a_{n}\sum\limits_{i}\gamma _{i}^{n}\beta _{i}^{n}x_{i}\right) ^{M+1}-1}{
\left( a_{n}\sum\limits_{i}\beta _{i}^{n}x_{i}\right) ^{M+1}-1}\right]
\left( \frac{a_{n}\sum\limits_{i}\beta _{i}^{n}x_{i}-1}{c_{n}a_{n}\sum
\limits_{i}\gamma _{i}^{n}\beta _{i}^{n}x_{i}-1}\right) -x_{n}.
\end{equation}
The advantage of this model consists in the fact that at each iteration
step, in the algorithm used to solve the system of differential equations,
one has to calculate only the first rank interaction tensors, $
\sum\limits_{i}\beta _{i}^{n}x_{i}$ and $\sum\limits_{i}\gamma _{i}^{n}\beta
_{i}^{n}x_{i}$, even though the expansion is considered up to the maximum
rank $M$.

We have performed a numerical experiment in which all the parameters and the
initial states are set as before for $M=1,2$. Also, the coefficients
regulating the monomer interactions $\beta _{i}^{n},\gamma _{i}^{n}\in
\{0,1\}$, $(i,n=1,...,N)$ are generated using the probabilities $p$, $q$, as
in the first numerical experiment, performed for $M=1$. Thus, the variable
parameters are $p$, $q$ and the maximum rank $M$ of tensorial expansion
(which corresponds to the maximum length of multimers mediating the
interaction among genes). Depending on the values of all these parameters,
the system exhibits different types of behavior. However, the most important
parameter seems to be $M$. For low values $1\leq M\leq 5$ the density of
steady states in the solution space is very high and therefore the system
converges most of the time to a steady state (Fig. 4 (a)). By increasing $M$
to $5\leq M\leq 8$ the density of steady states decreases and the typical
behavior becomes oscillatory (Fig. 4 (b)). For larger values of $M\geq 8$,
the dynamics becomes more complex, exhibiting chaotic behavior (Fig. 4 (c)).
The numerical results for this experiment show that there is a smooth
transition between order and chaos as the parameter $M$ increases. This is a
consequence of the fact that by increasing $M$ one actually increases the
cooperativity among the transcription factors and implicitly the
nonlinearity of the system.

\section{Conclusion}

We have discussed a mean-field model of genetic regulatory networks, in
which the regulatory interactions are described by differential equations
corresponding to the chemical equations considered in the network. We have
shown that, depending on the set of regulatory parameters, the model
exhibits differing behaviors corresponding to ordered and chaotic dynamics.
This result gives some quantitative support to the earlier qualitative
results obtained for random Boolean networks [2-6]. However, contrary to
Boolean networks, by increasing the number of interactions per gene, our
model acquires stability. This is an important issue which we would like to
address here and in the future. We believe that the stability of the system
occurs from the intrinsic construction of the activation/repression ratio.
This ratio corresponds to an average interaction, which takes into account
all the repression and activation mechanisms acting on any one gene in the
network. Also, according to the central limit theorem, the variance of the
sums, defining the numerator and denominator of this ratio, decreases by
considering more terms (interactions). These mechanisms create stability in
the system by producing a contraction which keeps the solution bounded. So,
by increasing the number of interactions, the system becomes more stable.
For specific sets of parameters the solution is not only bounded but it also
corresponds to ordered dynamics (steady states or oscillations). By changing
the set of parameters, this contraction becomes loose enough that the
solution becomes chaotic. In this case the system becomes ergodic and it is
able to explore large regions in the solution space. An important role in
generating oscillatory and chaotic dynamics is played by the length of
transcription factor multimers mediating the interaction among genes. Our
analysis has shown that protein multimerization is a necessary condition for
the discussed mean-field model to generate oscillatory and chaotic dynamics.

In summary, we have introduced and provided an initial analysis of a
mean-field model for a class of reasonably realistic chemical equations
modeling genetic regulatory networks. We presume a critical phase transition
occurs as the system goes from order to chaos. This is important because
recent evidence tentatively suggests that yeast cells are critical [23].
Indeed, it has been a long standing hypothesis that cells are critical or
slightly subcritical to withstand noise [24]. Thus, our results give
preliminary support to this hypothesis. It remains for future work to
explore in more detail how networks with diverse topologies but similar
kinetics behave. If it proves true that biological cells are critical or
near critical, our model should be of use in exploring the combinations of
network topologies, motifs and kinetic rules that can correspond to such
critical behavior.

\section*{Acknowledgements}

This research was supported by iCORE under Grant No. RT732223.

\newpage

\clearpage
\begin{figure}
\centering
\includegraphics[width=12cm]{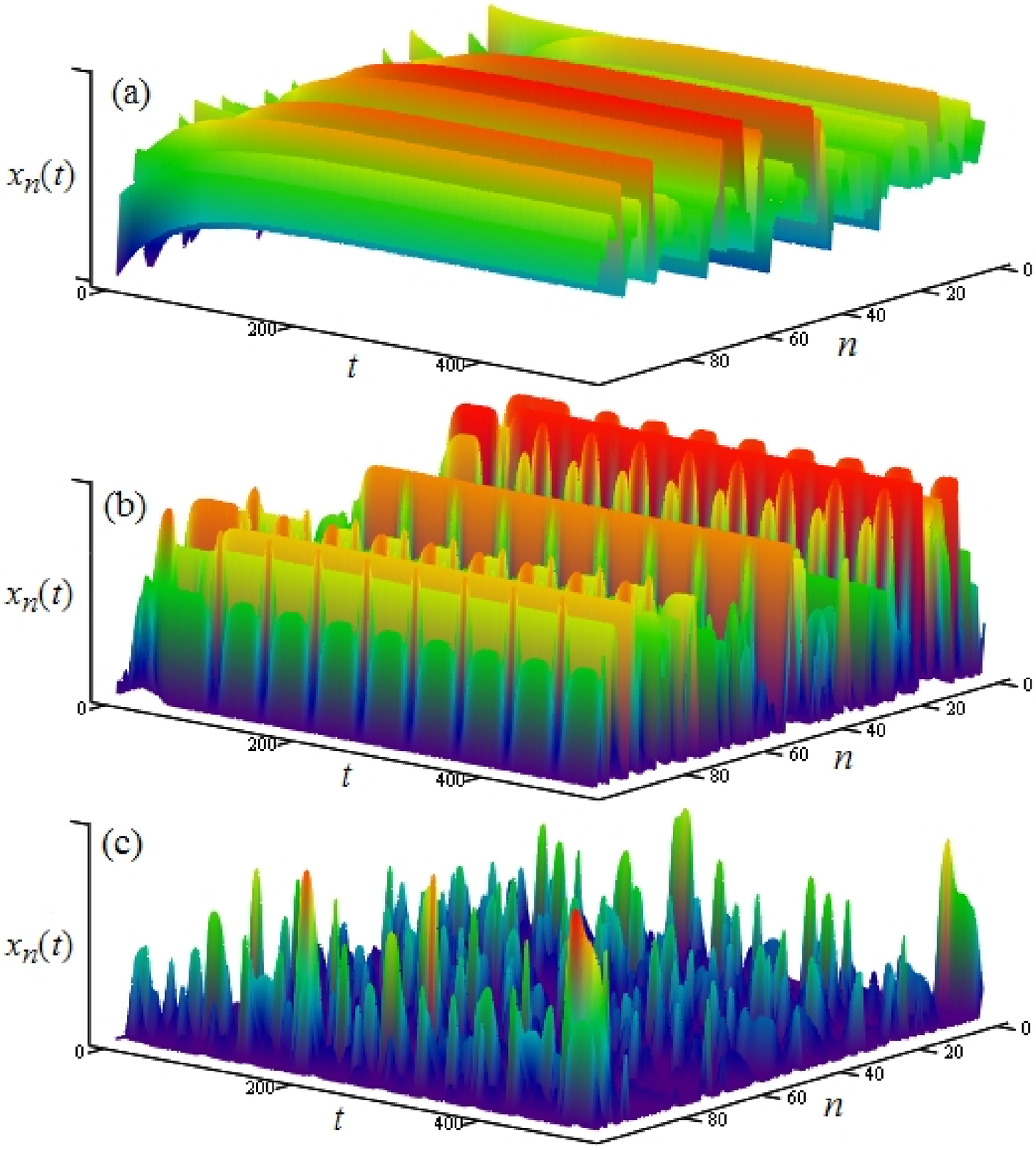}
\caption{\label{Fig1} Example of typical solutions
obtained for M=2: (a) steady state; (b) oscillations; (c) chaos. Here, $x_{n}(t)$ 
is the reduced concentration of protein monomer (transcription
factor) $n$ as a function of time $t$. The total number of genes in the
network is $N=100$.}
\end{figure}

\clearpage
\begin{figure}
\centering
\includegraphics[width=6cm]{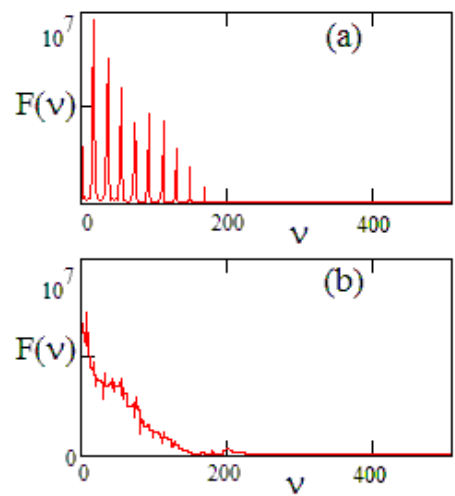}
\caption{\label{Fig2} Example of Fourier power
spectra obtained for single trajectories $x_{n}(t)$: (a) oscillations; (b) chaos.}
\end{figure}

\clearpage
\begin{figure}
\centering
\includegraphics[width=8cm]{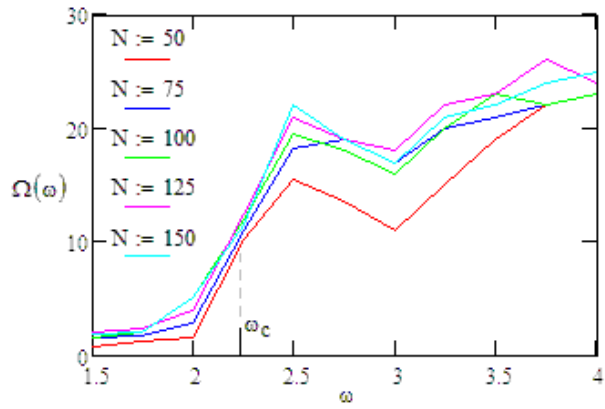}
\caption{\label{Fig3} The transition between a phase with
high density of ordinary steady state solutions ($\Omega \sim 0$) and a
phase with low density of steady state solutions ($\Omega \gg 0$).}
\end{figure}

\clearpage
\begin{figure}
\centering
\includegraphics[width=12cm]{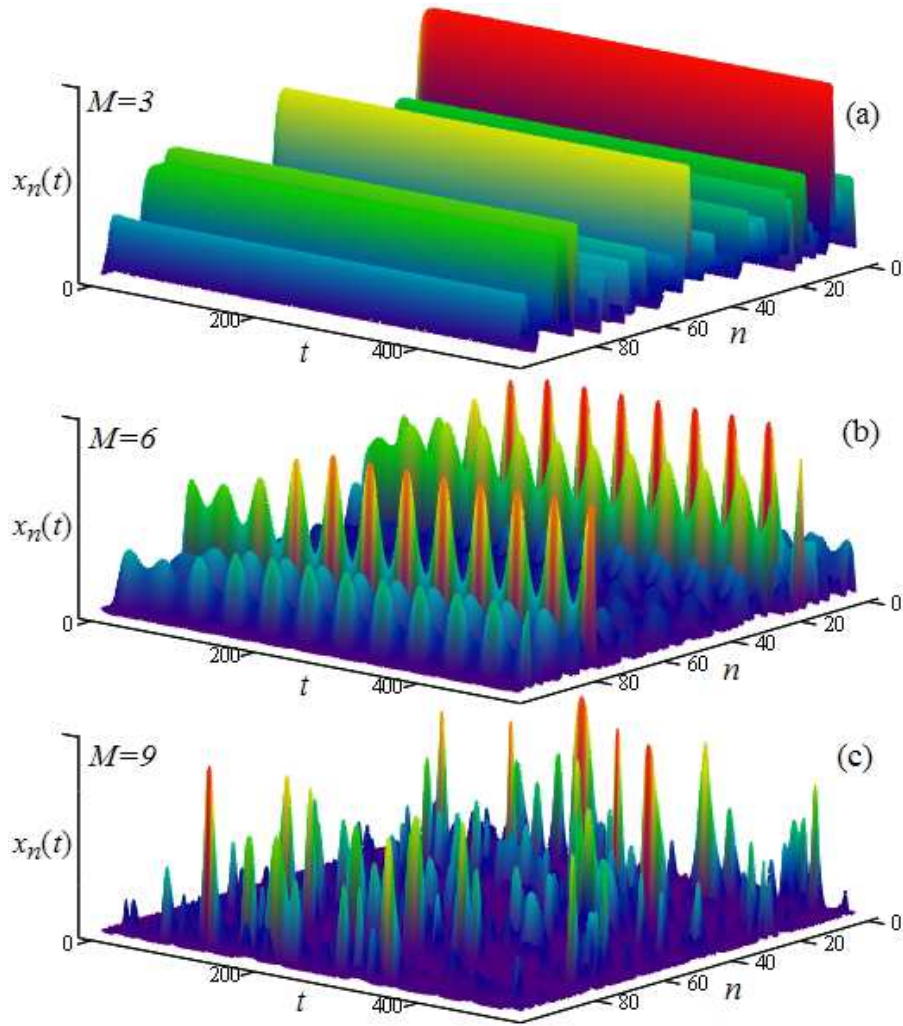}
\caption{\label{Fig4} Example of typical solutions
obtained for $M\geq 3$: (a) steady states; (b) oscillations; (c) chaos.
Here, $x_{n}(t)$ is the reduced concentration of protein monomer
(transcription factor) $n$ as a function of time $t$. The total number of
genes in the network is $N=100$. The parameter $M$ is the maximum length of
multimers mediating the interaction among genes.}
\end{figure}

\end{document}